\documentclass[aps,prl,reprint,fleqn]{revtex4-2}

\draft 
\usepackage{amssymb,amsmath}
\usepackage{graphicx}
\usepackage{dcolumn}
\usepackage{bm}
\usepackage{color}
\usepackage{float}
\usepackage{hyperref}
\usepackage[version=4]{mhchem}
\usepackage{relsize}
\usepackage{comment}
\usepackage{media9}

\definecolor{dark_green}{rgb}{0.0, 0.5, 0.0}

\newcommand{\OK}{\color{black}}

\begin{document}

\title{Neural Echo State Network using oscillations
of gas bubbles in water}

\author{Ivan S.~Maksymov}
\email{imaksymov@swin.edu.au}
\affiliation{Optical Sciences Centre, Swinburne University of
  Technology, Hawthorn, VIC 3122, Australia\looseness=-1} 

\author{Andrey Pototsky and Sergey A.~Suslov}
\affiliation{Department of Mathematics, Swinburne University of
  Technology, Hawthorn, Victoria 3122, Australia\looseness=-1} 


\begin{abstract}
In the framework of physical reservoir computing (RC),
machine learning algorithms designed for digital computers
are executed using analog computer-like nonlinear physical
systems that can provide energy-efficient computational
power for predicting time-dependent quantities that can be found
using nonlinear differential equations. We suggest a bubble-based
RC (BRC) system that combines the nonlinearity of an acoustic
response of a cluster of oscillating gas bubbles in water with
a standard Echo State Network (ESN) algorithm that is well-suited
to forecast chaotic time series. We confirm the plausibility of
the BRC system by numerically demonstrating its ability to
forecast {\OK certain chaotic time series similarly to 
or even more accurately than ESN}.
\end{abstract}

\maketitle 

Forecasting the time evolution of dynamical systems is important for
understanding many natural phenomena such as the behaviour of 
living organisms and the variation of the Earth's climate, for
predicting stock markets and for controlling autonomous  
vehicles \cite{Sma05}. However, nonlinearity of such systems
considerably complicates the task of prediction, which
forces the modern machine learning (ML) algorithms to rely
on longer observation times. Substantial computational
resources are needed to process such big data sets. 

\begin{figure}[t]
  \centerline{\includegraphics[width=8.5cm]{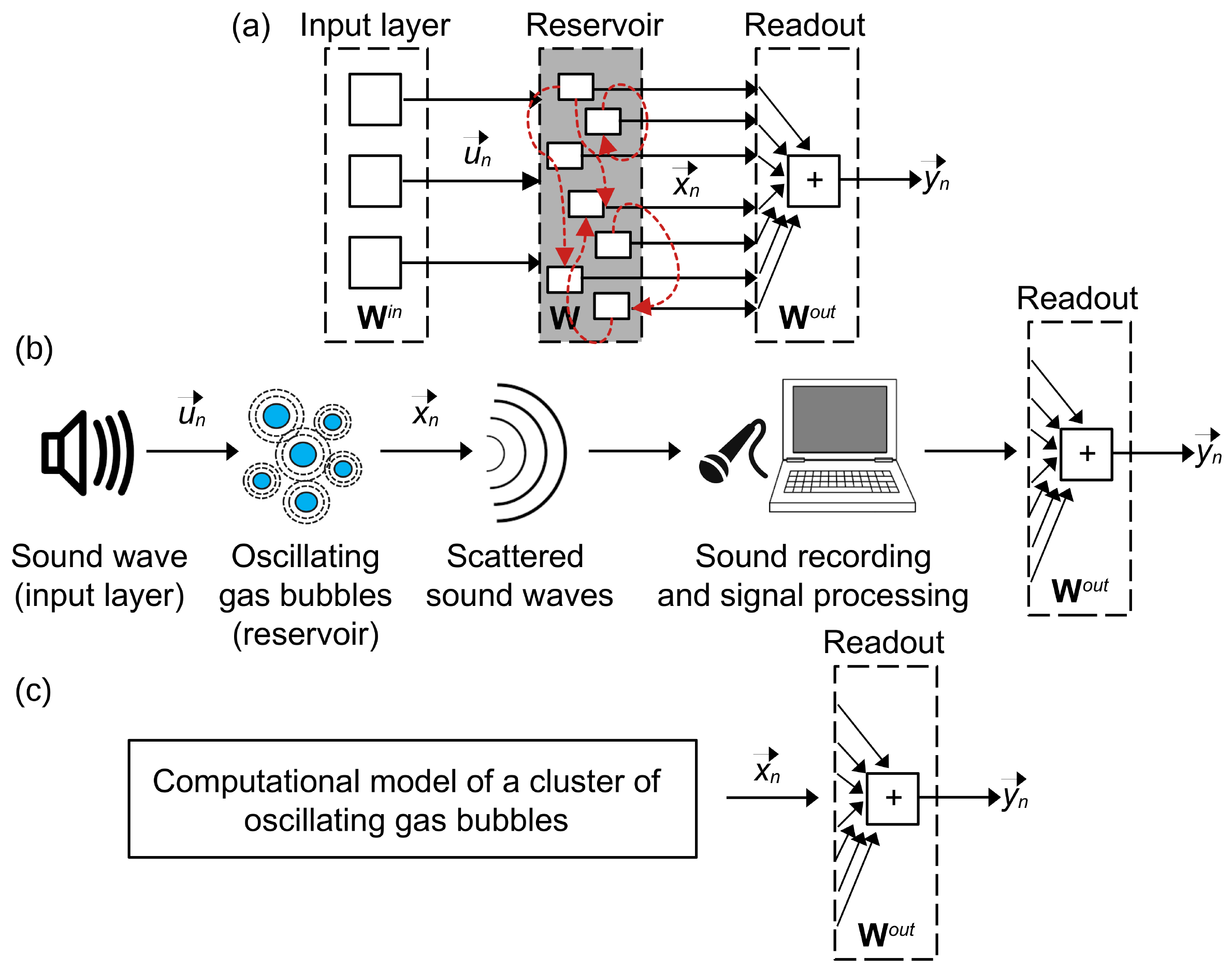}}
  \caption{(a)~ESN algorithm: the reservoir is a network of
    interconnected dynamical components (shown by dashed arrows); only
    the linear readout is trained to produce the output. (b)~The
    approach proposed in this work: complex interactions between the
    oscillating gas bubbles in water play the role of a reservoir, but
    the readout is trained using the ESN algorithm. (c)~Schematic of
    simulations used to demonstrate the plausibility of the
    BRC system. \label{fig:Fig1}} 
\end{figure}

Reservoir computing (RC) \cite{Luk09, Ver07, Gau21, Nak21} and
its foundation concepts of Echo State Networks (ESNs)
\cite{Jae05, Luk09} and Liquid State Machines (LSMs)
\cite{Maa02, Maa04} underpin an emergent approach to ML 
that is especially well-suited for forecasting the response of
nonlinear dynamical systems that exhibit chaotic or complex
spatiotemporal behaviour \cite{Jae05, Pat17, Pat18, Cha20, Gau21},
the problem that is a difficult to resolve using traditional 
ML algorithms \cite{Che20}. In an RC system
[Fig.~\ref{fig:Fig1}(a)], an artificial neural network
is structured as a combination of a fast output layer and a much
slower body of the network called the reservoir \cite{Sch05}. Such an 
arrangement helps resolving especially challenging forecasting
problems including free-running generation of time series,
where a trained RC system is presented with previously unseen
data and is tasked with making a forecast using its own output
from the previous time steps \cite{Luk12, Luk21, Gau21}.

The standard ESN algorithm laying a foundation of RC uses
the following nonlinear update equation
\cite{Jae05, Luk09, Luk12}:
\begin{eqnarray}
  \vec{x}_{n} = (1-\alpha)\vec{x}_{n-1}+ 
  \alpha\tanh({\bf W}^{in}\vec{u}_{n}+{\bf W}\vec{x}_{n-1})\,,
  \label{eq:RC1}
\end{eqnarray}
where $n$ is the index denoting entries corresponding to
equally-spaced discrete time instances $t_n$, $\vec{u}_n$ is the
vector of $N_u$ input values, $\vec{x}_n$ is a vector of $N_x$
neural activations of the reservoir, the operator $\tanh(\cdot)$ applied
element-wise to its arguments is a typical sigmoid activation
function used in the nonlinear model of a neuron \cite{Hay98},
${\bf W}^{in}$ is the input matrix consisting of $N_x \times N_u$
elements, ${\bf W}$ is the recurrent weight matrix containing
$N_x \times N_x$ elements and $\alpha \in (0, 1]$ is the leaking
rate that controls the update speed of the reservoir's
temporal dynamics. 

To train the linear readout of ESN one calculates the output
weights ${\bf W}^{out}$ by solving a system of linear
equations ${\bf Y}^{target} = {\bf W}^{out}{\bf X}$, where the
state matrix ${\bf X}$ and the target matrix ${\bf Y}^{target}$
are constructed using, respectively, $\vec{x}_n$ and the vector
of target outputs $\vec{y}_n^{target}$ as columns for each time
instant $t_n$. The solution is often obtained in the form 
${\bf W}^{out} = {\bf Y}^{target} {\bf X^\top} ({\bf X}{\bf X^\top} +
\beta {\bf I})^{-1}$, where ${\bf I}$ is the identity matrix,
$\beta=10^{-8}$ is a regularisation coefficient and ${\bf X^\top}$ is
the transpose of ${\bf X}$ \cite{Luk12}. Then, one uses the
trained ESN, solves Eq.~(\ref{eq:RC1}) for new input data
$\vec{u}_n$ and computes the output vector
$\vec{y}_n={\bf W}^{out}\vec{x}_n$ (in our case a more common
form $\vec{y}_n={\bf W}^{out}[1;\vec{u}_n;\vec{x}_n]$ optimised
to forecast complex nonlinear time series using a constant bias and 
the concatenation $[\vec{u}_n;\vec{x}_n]$ \cite{Luk09, Luk12} produced
qualitatively similar results). We note that ESN needs to know the
target data only when it is trained since for forecasting it
uses its own output from the previous time step, i.e.~$\vec{x}_{n}$
is calculated using Eq.~(\ref{eq:RC1}) with
$\vec{u}_{n} = \vec{y}_{n-1}$. However, the target data may still
be needed to assess the accuracy of the forecast made by ESN. 

The performance tests of ESN are conducted using 
  target chaotic Mackey-Glass time series (MGTS) \cite{Luk12} that 
  are produced by the delay differential equation \cite{Mac77}
\begin{eqnarray} 
  \dot{x}_{_{MG}}(t) 
  &=&\beta_{_{MG}}\frac{x_{_{MG}}(\tau_{_{MG}}-t)}
      {1+x_{_{MG}}^{q}(\tau_{_{MG}}-t)}-\gamma_{_{MG}}x_{_{MG}}(t)\,,
  \label{eq:MG}
\end{eqnarray}
where one typically chooses $\tau_{_{MG}}=17$ and sets
$q=10$, $\beta_{_{MG}}=0.2$ and $\gamma_{_{MG}}=0.1$ \cite{Luk12}.
MGTS with these parameters is uniquely suited for a demonstration
of the abilities of ESN to forecast chaotic time series \cite{Jae04}.

However, ESN is essentially a program for a digital computer and its
ability to forecast is limited by the available computational
resources. Therefore, it has been suggested that the ESN algorithm can
be implemented using certain non-digital nonlinear physical systems
\cite{Tan19, Nak20, Nak21}, which at the conceptual model level means
that Eq.~(\ref{eq:RC1}) is replaced with the respective nonlinear
differential equation describing dynamics of a particular physical
system. Similarly to analogue computers that can solve certain
problems more efficiently than their digital counterparts
\cite{Hyn70, Cow05_paper, Sor15, Zan18}, physical RC systems
may provide energy efficiency in practical situations,
where the relationship between time-dependent physical
quantities that needs to be predicted can be expressed
using solutions of nonlinear differential equations. A
number of physical RC networks have been demonstrated using spintronic
systems \cite{Fur18, Rio19, Wat20}, liquids \cite{Fer03, Nak15},
quantum ensembles \cite{Fuj17} and electronic \cite{Nak21}, photonic
\cite{Nak21, Sor19} and mechanical devices \cite{Cou17, Nak21}.

Here, we propose [Fig.~\ref{fig:Fig1}(b)] and computationally
validate [Fig.~\ref{fig:Fig1}(c)] an approach to RC, where the
nonlinear dynamics of a reservoir is represented by weakly
nonlinear oscillations of a cluster of gas bubbles in water
driven by an acoustic pressure wave \cite{Lau10}. Previously,
we demonstrated that in a cluster of mm-sized bubbles with
randomly-chosen equilibrium radii and initial spatial positions
each bubble emits a unique acoustic signal that reflects a complex
nature of its interaction with the neighbouring bubbles \cite{Tony21}.
Furthermore, we suggested an acoustic frequency comb technique
that can be used to reliably detect such signals \cite{Mak21, Tony21}.
Thus, a cluster consisting of $N_b$ randomly sized and positioned
bubbles can be used as a reservoir network of $N_b\times N_b$
random connections, where the acoustic response of individual bubbles
can serve as a physical counterpart of the neural activation states
of ESN \cite{Jae05, Luk09}. 

We replace Eq.~(\ref{eq:RC1}) by Rayleigh-Plesset equation of
nonlinear dynamics of spherical gas bubble oscillations
\cite{Lau10, Tony21} in a cluster consisting of $N_b$ bubbles
not undergoing translational motion \cite{Lau10, Tony21}:  
\begin{eqnarray} 
  R_p\ddot{R}_p
  &+&\frac{3}{2}{\dot{R}_p}^2 = \frac{1}{\rho}\left[P_p
      -P_\infty(t)\right]-P_{sp}\,,\label{eq:eq1}
\end{eqnarray}
where overdots denote differentiation with respect to time and for the
$p$th bubble in the cluster
\begin{eqnarray}
  P_p = \left(P_0-P_v+\frac{2\sigma}{R_{p0}}\right)
      \left(\frac{R_{p0}}{R_p}\right)^{3\kappa} 
  -\frac{4\mu}{R_p}\dot{R}_p-\frac{2\sigma}{R_p}\,.\label{eq:eq2}
\end{eqnarray}
The term accounting for the pressure acting on of the $p$th bubble
due to scattering of the incoming pressure wave by the neighbouring
bubbles in a cluster is given by 
\begin{equation}
  \label{eq:eq3}
  P_{sp}=\sum_{l=1, l\neq p}^{N_b}\dfrac{1}{d_{nl}}\left( R_l^2\ddot{R}_l
    + 2R_l{\dot{R}_l}^2 \right)\,,
\end{equation}
where $d_{pl}$ is the inter-bubble distance and parameters $R_{p0}$ and
$R_p(t)$ are the equilibrium and instantaneous radii of the $p$th bubble
in the cluster. The term $P_\infty(t)=P_0-P_v+\alpha_s u_s(t)$
represents the time-dependent pressure in the liquid far from the bubble,
where $\alpha_s$ and $u_s(t)$ are the amplitude and temporal profile of
the sound wave driving oscillations of the bubbles. The acoustic power
scattered by the $p$th bubble in the cluster in the far-field zone 
at the distance $h$ is \cite{Tony21}
\begin{equation}
  \label{eq:eq4}
  P_{scat}(R_p,t) = \frac{\rho R_p}{h}\left(R_p\ddot{R}_p+2\dot{R}_p^2\right)\,. 
\end{equation}

To incorporate Eq.~(\ref{eq:eq1}) into the linear readout of ESN,
we sample $P_{scat}(R_p,t)$ and $u_s(t)$ at equidistant time
instances and obtain their discrete analogs that we treat as
the vectors of neural activations $\vec{x}_n$ and of input values
$\vec{u}_n$ of ESN, respectively. To train the BRC system, we
use MGTS $x_{_{MG}}(t)$ obtained from Eq.~(\ref{eq:MG}) as the
driving sound signal $u_s(t)$. However, when the BRC system makes
a forecast, it does not know the target series because $u_s(t)$
is defined by the discrete network output $\vec{y}_n$.       

The signal amplitude $u_s(t)$ is chosen to be small
($\alpha_s = 0.1-1$\,kPa, i.e.~$\alpha_s\ll P_0$) so that 
both the nonlinearity of bubble oscillations and Bjerknes forces 
acting between bubbles in the cluster \cite{Lau10} remain weak.
Such physical conditions allow a cluster of mm-sized bubbles
to remain stable over a time sufficient to train and exploit
the BRC system before the configuration of the cluster
changes due to translational motion of bubbles \cite{Tony21}.
The cluster stability is important because the topology of a
reservoir must not change during its training and use
\cite{Luk12}. The operation of the BRC system in a weakly
nonlinear regime also helps satisfying the echo state condition
that implies that dynamics of the neural activations
$\vec{x}_n$ is uniquely defined by a given input signal
$\vec{u}_n$ \cite{Jae05}. Physically, this means that the phase
space of Eq.~(\ref{eq:eq1}) does not contain multiple
periodic or chaotic attractors or fixed points.

\begin{figure}[t]
  \centerline{
    \includegraphics[width=5cm]{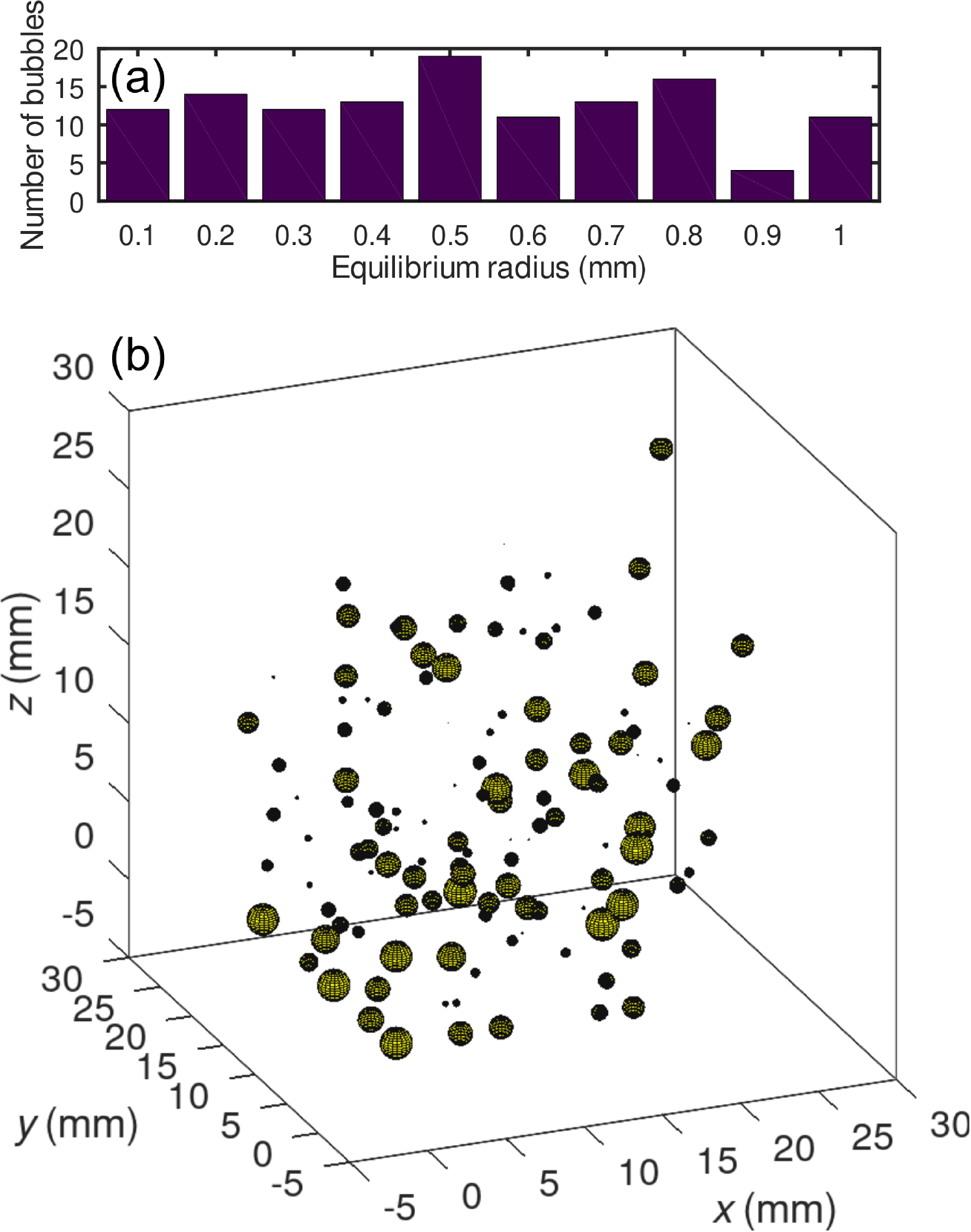}}
  \caption{(a)~Size and (b)~spatial distribution of bubbles in a
    representative cluster used as the physical reservoir. 
    \label{fig:Fig2}}
\end{figure}

For our simulations, we generate a cluster consisting of 125 bubbles
with equilibrium radii randomly chosen in the 0.1 to 1\,mm range
(Fig.~\ref{fig:Fig2}). We use the following model parameters
corresponding to water at $20^\circ$\,C: $\mu=10^{-3}$\,kg\,m/s, 
$\sigma=7.25\times10^{-2}$\,N/m, $\rho=10^3$\,kg/m$^3$,
$P_v=2330$\,Pa, $P_0=10^5$\,Pa and $\kappa=4/3$ \cite{Tony21}.
Relevant computational details can also be found in \cite{Tony21}.

The dynamics of a cluster of oscillating bubbles also has to
match the speed of temporal evolution of the training time series.
In ESN, this is achieved using the leaking rate $\alpha$
in Eq.~(\ref{eq:RC1}) \cite{Luk12}. In our BRC system, the
dynamics of the network is controlled by ensuring that the
frequency of the major peak $f_{_{MG}}$ in the Fourier
spectrum of the time series $x_{_{MG}}(t)$ is close to the
frequency of natural oscillations of individual bubbles
with the most representative equilibrium radius in
Fig.~\ref{fig:Fig2}(a) (using the well-known Minnaert formula
\cite{Lau10} we obtain $f_{_{MG}} \approx 6.5$\,kHz). To tune
$f_{_{MG}}$ it is convenient to change the time scale of
previously tabulated $x_{_{MG}}(t)$, for example, by using
an auxiliary discrete time instant in a computer program that
solves Eq.~(\ref{eq:eq1}). Once the trained network has
produced an output signal, the original time scale is restored.

The advantage of such a rescaling procedure is that a cluster of
mm-sized bubbles can be used to forecast time series with disparate
timescales. Indeed, the dynamics of the BRC system could also be accelerated
by decreasing the equilibrium radius of bubbles (and thus increasing
their natural frequency). However, generation of microscopic bubbles
requires special techniques \cite{Che09}. Besides, microscopic bubbles
are effectively stiffer than mm-sized ones \cite{Lau10} so that
measurements involving them require special high-frequency and
high-power ultrasonic equipment compared with a technically simple
acoustic setup sufficient for studying mm-sized bubbles \cite{Mak21}. 

\begin{figure*}[t]
  \centerline{
    \includegraphics[width=10.0cm]{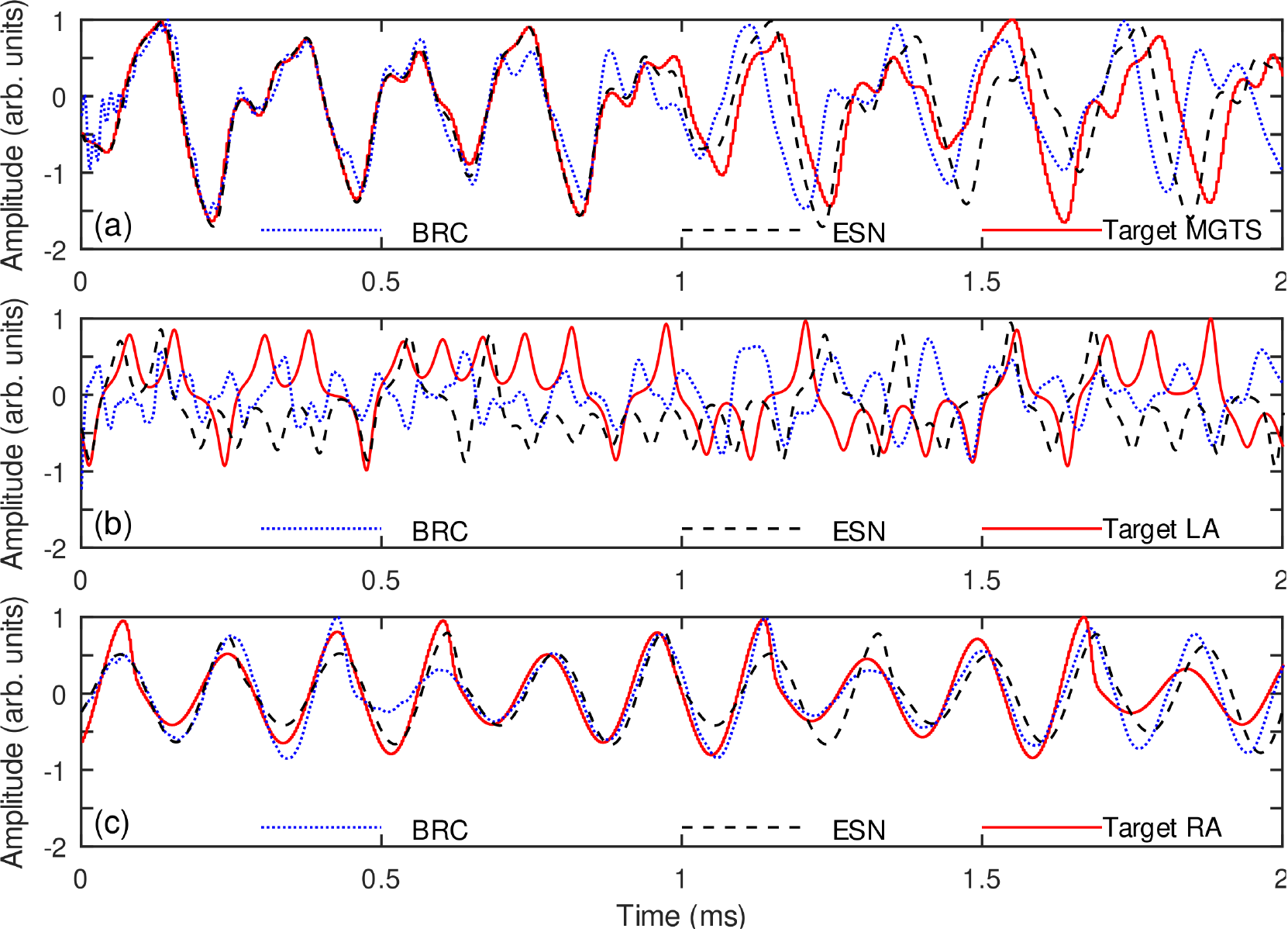}}
  \caption{\OK The BRC (the dotted lines) and ESN (the dashed lines)
  forecasts compared with the target (a)~MGTS, (b)~LA and (c)~RA (the
  solid lines). The target data, which are plotted here for reference
  only, are not used by the RC systems. In ESN, the leaking rate
  $\alpha$ equals 0.3, 0.1 and 0.01 for MGTS, LA and RA, respectively.}
  \label{fig:Fig3}
\end{figure*}

In Fig.~\ref{fig:Fig3}(a), we demonstrate the ability of a
trained BRC system to forecast MGTS. We also compare the accuracy of its
forecast with that of ESN with $N_x=125$, which is equivalent to the
size of the BRC reservoir. {\OK In the time interval $0-1$\,ms
the BRC system correctly reproduces both the pattern and the timeline of MGTS.
The mean-square error (MSE)---a standard measure of the accuracy of ESNs
\cite{Luk12}---is approximately $5\times 10^{-2}$ for the BRC system,
which is two orders of magnitude larger than that for ESN thereby
indicating that in a short-term perspective ESN performs better.
However, over the full test interval $0-2$\,ms MSE of both RC systems
is approximately $0.5\times 10^{-2}$, which indicates that the long-term
behaviour of the BRC system may be closer to the target than that of ESN.}

Nevertheless, the observed behaviour of ESN is widely regarded as a
positive outcome for chaotic systems with limited availability of
information \cite{Jae05, Luk09, Gau21}. Indeed, many other competitive
neural network architectures either fail to deliver similar results
using small data sets or can produce acceptable results only using
much bigger data sets (and therefore requiring substantial
computational resources) \cite{Sch05, Che20, Luk09}. These features 
make ESN the best-in-class ML algorithm designed to forecast highly
nonlinear and chaotic time series \cite{Gau21}. Given this, {\OK the
ability of a much simpler BRC system to predict MGTS similarly to, and
in some aspects matching the target better than ESN is a significant
result. While without any optimisation attempted so far the
  accuracy of the BRC system may be not as high as that of ESN in a
  short term, it can be advantageously used in applications tolerating
  a lower accuracy of the forecast \cite{Jun14} or if a long-term
  reliability of a forecast is the priority}.

Results quantitatively similar to those in Fig.~\ref{fig:Fig3}(a)
were obtained for other clusters that were randomly
generated using the same set of equilibrium bubble radii.
Significantly, all forecast time series had a phase lag with
respect to the target signal. The existence of a phase shift
between the driving sound pressure and acoustic power
scattered by a single gas bubble is a well-established fact 
\cite{Lau10}. This effect is also present in the bubble-cluster
reservoir, where bubble oscillations are driven by a continuous
acoustic signal produced using the discrete output of the network.
The phase lag was essentially the same for all random cluster
configuration of similarly sized bubbles. This is because the
bubble response delay is caused by the inertia of liquid
surrounding them that depends only on their sizes. Subsequently,
this phase lag was removed from all relevant results during
post-processing.

{\OK 
To forecast time series that exhibit a more chaotic behaviour 
than MGTS, ESN require a larger reservoir and a delicate
case-by-case tuning of its algorithmic parameters\cite{Luk12}.
On the other hand, the BRC system could solve specific classes
of problems more efficiently than ESN without the need to
modify the configuration of the gas bubble cluster. In
particular, similarly to certain analogue computational
systems \cite{Vee47, Hyn70}, the use of our BRC system 
could help avoiding limitations imposed by time-step
discretisation needed to numerically solve differential
equations arising in practice \cite{Cow05_paper, Sor15, Zan18}.}
  
{\OK
To verify this assertion, we compare the performance and
implementation cost of ESN and the BRC system to forecast
the behaviour of Lorenz (LA) \cite{Lor63} and R{\"o}ssler (RA)
\cite{Ros76} attractors. We keep the same parameters for both
RC systems as in the tests with MGTS, but allow variations of
the leaking rate $\alpha$ of ESN. The implementation details
and discussion of the system performance can be found in Supplemental
Material. As seen in Fig.~\ref{fig:Fig3}(b), neither BRC system nor
ESN can follow the long-term behaviour of LA. ESN may be able to mimic
the behaviour of LA somewhat better initially but overall it suffers
from an apparent negative bias underpredicting the LA output over
the most of the test interval. In contrast, the BRC system produces
values that are on average much closer to the target but it misses
some of fine details of the LA behaviour. Indeed, in the time
interval $0-2$\,ms the BRC system has MSE=0.33 compared with
0.55 for ESN, which speaks in favour of the BRC system. At the same
time, the BRC system has a clear advantage in terms of the
implementation cost. For example, to obtain the ESN prediction of LA
presented in Fig.~\ref{fig:Fig3}(b) a lengthy procedure of tuning the
value of the leaking rate $\alpha$ had to be followed to avoid
numerical instabilities caused by artefacts of time-step
discretisation of Eq.~(\ref{eq:RC1}) and by the fact that the size of
the chosen reservoir had to be kept small to enable a meaningful
comparison (in general, to produce a qualitatively accurate
  output ESN requires a reservoir with a much larger size than that
  used here).

Whereas significant tuning was also needed to enable ESN to
forecast RA, the BRC system could forecast RA without any
optimisation [Fig.~\ref{fig:Fig3}(c)]. Significantly, MSE
of the BRC system for the interval $0-2$\,ms is approximately 0.06
compared with 0.31 for ESN. This result strongly speaks
in favour of the proposition that the BRC system could
outperform ESN in some practical situations. We established that
the BRC system can predict RA because, similarly to MGTS,
the Fourier spectrum of RA has well-defined frequency peaks
(see Supplemental Material). As a result, the oscillating bubbles
of the BRC reservoir can match the frequencies of the peaks.
In contrast, the spectrum of LA is continuous and has no
  well-defined discrete peaks. This is the reason why the BRC system
  inherently possessing a discrete spectrum cannot forecast LA or
  any other continuous-spectrum signal accurately. Several potential
  approaches to resolving this challenge are discussed in Supplemental
  Material.

In conclusion, we have demonstrated through numerical simulations
that an RC system employing a cluster of oscillating gas bubbles
in water as the reservoir can forecast certain chaotic time series
similarly to ESN. Although the currently achievable accuracy of
the proposed RC system may be lower than that of highly
optimised ESN, it can be increased using, for example, novel
techniques of gas bubble cluster manipulation \cite{Mak21,
  Tony21}.  In certain practically important cases
(e.g.~when the spectrum of the target signal has well-defined
peaks) the size of the BRC reservoir required to achieve comparable
accuracy may be significantly smaller than that of ESN, and it may
require no or little tuning compared to that needed for ESN.
Moreover, since its prototype can be built using energy-efficient
integrated electronic circuits and piezoelectric transducers,
BRC holds the promise of being less expensive to build and at the same
time more computationally and energy-efficient to run than an ESN
implemented on a workstation computer.}

\begin{acknowledgments}
  ISM acknowledges the support from the Australian Research Council
  via the Future Fellowship (FT180100343) program and 
  useful discussions with Professor Mikhail Kostylev (The University
    of Western Australia).
\end{acknowledgments}

\providecommand{\noopsort}[1]{}\providecommand{\singleletter}[1]{#1}%

\end{document}